\documentclass[twocolumn]{revtex4}

\usepackage{graphicx}
\usepackage{dcolumn}
\usepackage{amsmath}
\usepackage{amssymb}

\begin{document}

\title{The Stochastic Dynamics of Rectangular and V-shaped Atomic Force Microscope 
Cantilevers in a Viscous Fluid and Near a Solid Boundary}

\author{M. T. Clark}
 \email{clarkmt@vt.edu}
\author{M. R. Paul}
\affiliation{Department of Mechanical Engineering, Virginia
Polytechnic Institute and State University, Blacksburg, Virginia 24061}

\date{\today}

\begin{abstract}

Using a thermodynamic approach based upon the fluctuation-dissipation theorem 
we quantify the stochastic dynamics of rectangular and V-shaped microscale cantilevers 
immersed in a viscous fluid. We show that the stochastic cantilever dynamics as measured 
by the displacement of the cantilever tip or by the angle of the cantilever tip are different. We 
trace this difference to contributions from the higher modes of the cantilever.  We find that 
contributions from the higher modes are significant in the dynamics of the cantilever tip-angle.   
For the V-shaped cantilever the resulting flow field is three-dimensional and complex 
in contrast to what is found for a long and slender rectangular cantilever. Despite this complexity 
the stochastic dynamics can be predicted using a two-dimensional model with 
an appropriately chosen length scale. We also quantify the increased fluid dissipation that 
results as a V-shaped cantilever is brought near a solid planar boundary. 

\end{abstract}

\maketitle

\section{Introduction}
\label{section:introduction}

The stochastic dynamics of micron and nanoscale cantilevers immersed in 
a viscous fluid are of broad scientific and technological interest~\cite{butt:1995,ekinci:2005}. 
Of particular importance is the oscillating cantilever that is central to atomic force 
microscopy~\cite{binnig:1986,garcia:2002}. Significant theoretical progress has been 
made using simplified models in the limit of long and thin rectangular 
cantilevers~\cite{tuck:1969,sader:1998,clarke:2005}. In this case, a two-dimensional 
approximation is appropriate (therefore neglecting effects due to the 
tip of the cantilever) and has yielded important insights. However, it is not certain how well 
these approximations work for many situations of direct experimental 
interest. For example, a commonly used cantilever in atomic 
force microscopy is V-shaped and a theoretical description of the dynamics 
of these cantilevers in fluid is not available.

Furthermore, micron and nanoscale cantilevers are often used in close 
proximity to a solid boundary either by necessity or out of experimental 
interest. It is well known experimentally and theoretically 
that the presence of a solid boundary increases the fluid 
dissipation resulting in reduced quality factors and reduced resonant 
frequencies~\cite{benmouna:2002,nnebe:2004,clarke:2005,
clarke:2006,clarke:2006prs,green:2005,green:2005:jap,harrison:2007}. 
Again, theoretical descriptions are available in the limit of long and thin 
rectangular cantilevers and it is uncertain if these approaches can be 
applied to these more complex geometries.

In this paper we use a powerful thermodynamic approach to quantify the 
stochastic dynamics of cantilevers due to Brownian motion for
experimentally relevant geometries 
for the precise conditions of experiment including the presence of a planar 
boundary. Our results are valid for the precise three-dimensional geometry of 
interest and include a complete description of the fluid-solid interactions. Using 
these results we are able to compare with available  theory to yield further 
physical insights and to suggest simplified analytical approaches to describe 
the cantilever dynamics for these complex situations.
\begin{figure}[htb]
  \begin{center}
    \includegraphics[width=3in]{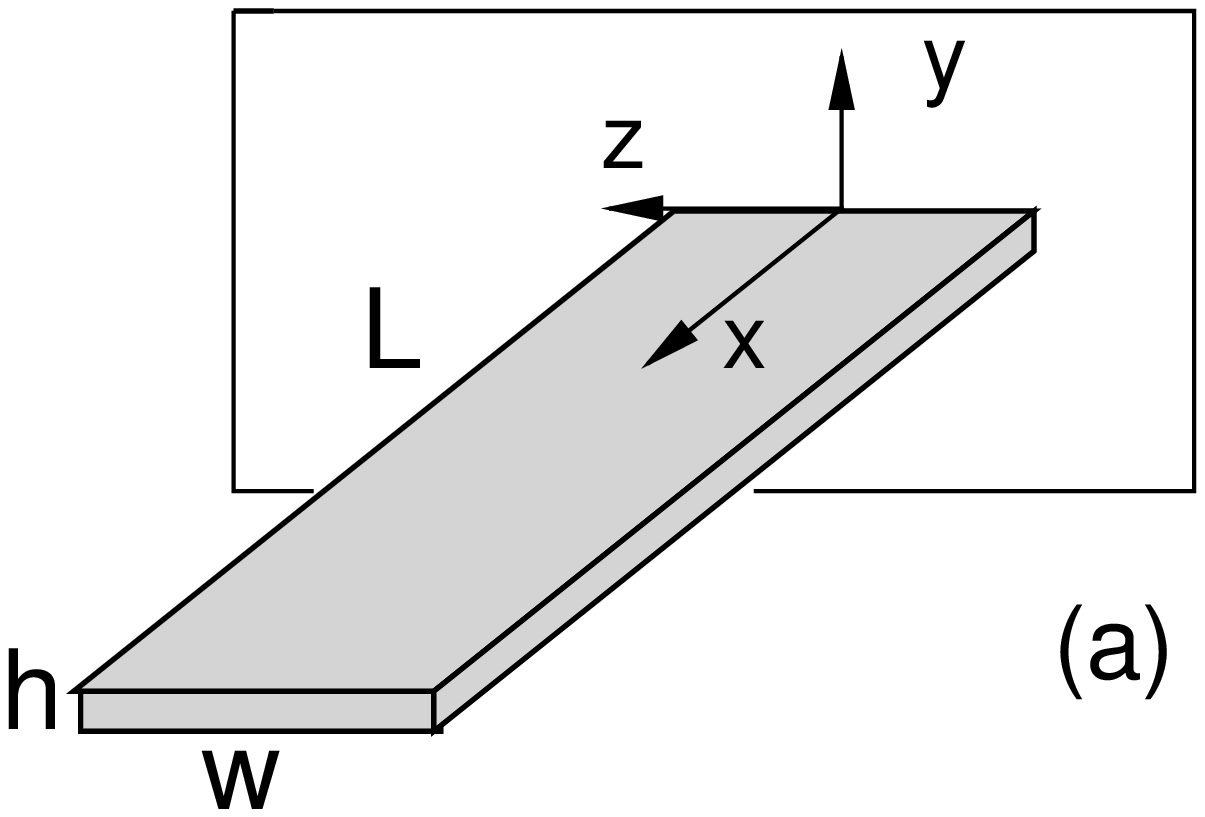}
    \includegraphics[height=2in]{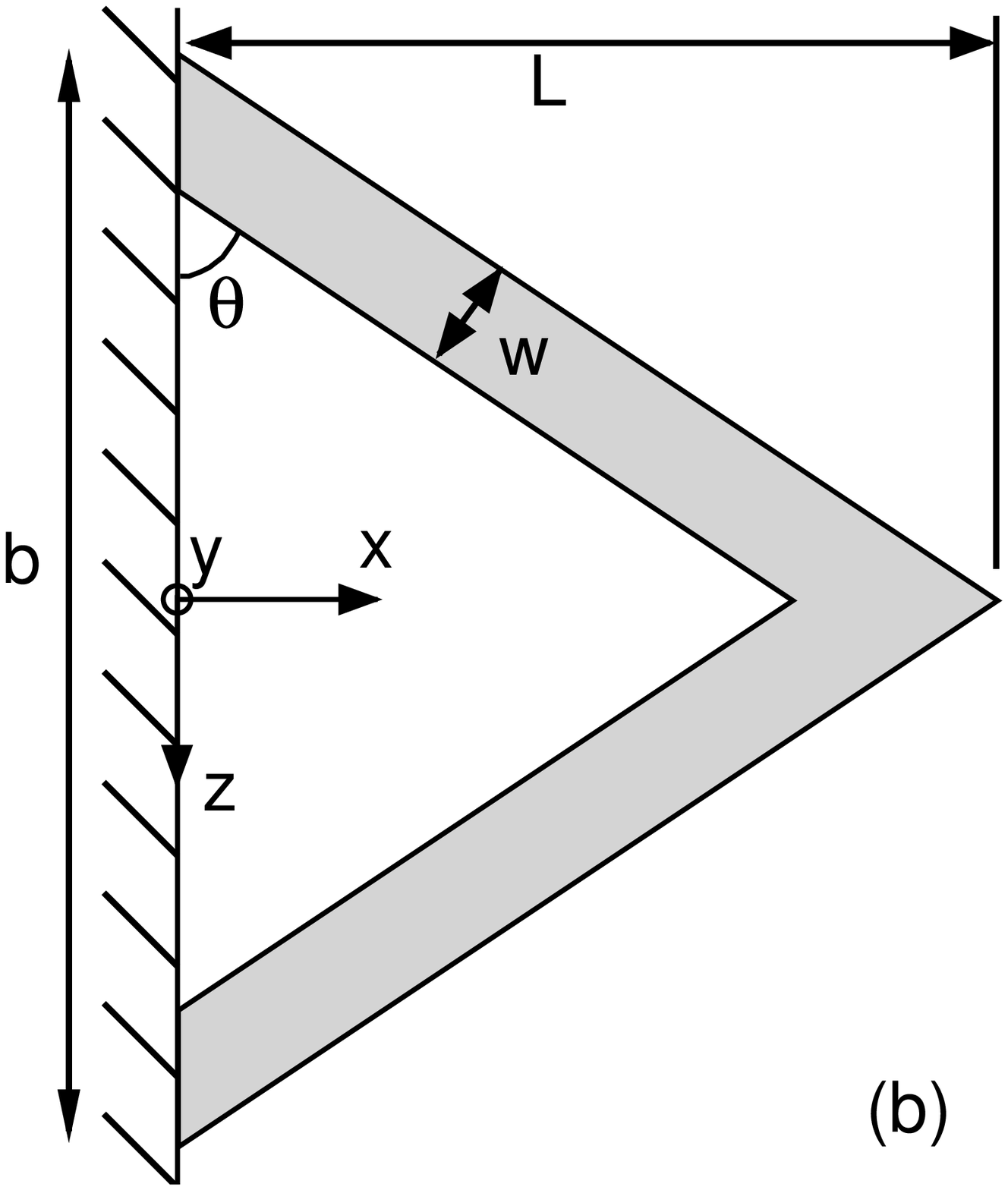}
  \end{center}
  \caption{Schematics of the two micron scale cantilever geometries considered (not drawn to scale). 
  Panel~(a),~A rectangular cantilever with aspect ratios $L/h = 98.5$, $w/h = 14.5$, and
  $L/w = 6.8$.  The cantilever is composed of silicon with density $\rho_c=$ 2329 kg/$\text{m}^3$  
  and Youngs Modulus $E =$ 174 GPa. Panel~(b),~A V-shaped cantilever with aspect 
  ratios $L/h = 233$, $w/h = 30$, and $L/w = 7.8$.  The total width between the two arms 
  normalized  by the width of a single arm is $b/w=10.36$. The cantilever planform is an equilateral 
  triangle with $\theta=\pi/3$. The cantilever is composed of silicon nitride with 
  $\rho_c=3100$kg/$\text{m}^3$ and $E=172$GPa. The specific dimensions for the rectangular 
  and V-shaped cantilever are  given in Table~\ref{table:geometry}.}
  \label{fig:beam}
\end{figure}
\section{Thermodynamic Approach -- Fluctuations from Dissipation}
\label{section:thermo}

The stochastic dynamics of micron and nanoscale cantilevers driven 
by thermal or Brownian motion can be quantified using strictly deterministic calculations. This is 
accomplished using the fluctuation-dissipation theorem since the cantilever remains near 
thermodynamic equilibrium~\cite{paul:2004,paul:2006}. We briefly review this approach for 
the case of determining the stochastic displacement of the cantilever tip and then extend it to the 
experimentally important case of determining the stochastic dynamics of the angle of the 
cantilever tip. 

The autocorrelation of equilibrium fluctuations in cantilever displacement can be 
determined from the deterministic response of the cantilever to the removal of a step force 
from the tip of the cantilever (i.e. a transverse point force removed from the distal end of the 
cantilever). If this force $f(t)$ is given by
\begin{equation}
f(t) =\left\{
\begin{array}
[c]{cc}
F_0 & \text{for }t<0\\
0 & \text{for }t\ge0,
\end{array}
\right. \label{eq:step_force}
\end{equation}
where $t$ is time and $F_0$ is the magnitude of the force, then the autocorrelation of the 
equilibrium fluctuations in the displacement of the cantilever tip is given directly by
\begin{equation}
\left< u_1(0) u_1(t) \right> = k_B T \frac{U_1(t)}{F_0},
\label{eq:autocorr_displacement}
\end{equation}
where $k_B$ is Boltzmann's constant, $T$ is the temperature, and $\left< \right>$ is an 
equilibrium ensemble average. In our notation  lower case letters represent 
stochastic variables ($u_1(t)$ is the stochastic displacement of the cantilever tip) and 
upper case letters represent deterministic variables ($U_1(t)$ represents the deterministic 
ring down of the cantilever tip due to the step force removal). The spectral properties of the 
stochastic dynamics are given by the Fourier transform of the autocorrelation. 

The thermodynamic approach is valid for any conjugate pair of variables~\cite{paul:2004}. 
For example, it is common in experiment to use optical techniques to measure the angle of the 
cantilever tip as a function of time~\cite{garcia:2002}. It has also been proposed to use 
piezoresistive techniques to measure voltage as a function of time~\cite{arlett:2006}. The 
thermodynamic approach remains valid for these  situations by choosing the correct conjugate 
pair of variables.

In this paper we also explore the stochastic dynamics of the angle of the cantilever tip. In this 
case, the angle of the cantilever tip is conjugate to a step point-torque applied to the 
cantilever tip. If this torque is given by
\begin{equation}
\tau(t)=\left\{
\begin{array}
[c]{cc}
\tau_0 & \text{for }t<0\\
0 & \text{for }t\ge0,
\end{array}
\right. \label{eq:step_torque}
\end{equation}
where $\tau_0$ is the magnitude of the step torque, then the autocorrelation of equilibrium 
fluctuations in cantilever tip-angle $\theta(t)$ is given by
\begin{equation}
\left< \theta_1(0) \theta_1(t) \right> = k_B T \frac{\Theta_1(t)}{\tau_0}.
\label{eq:autocorr_angle}
\end{equation}
Here $\Theta_1(t)$ represents the deterministic ring down, as measured by the tip-angle, 
resulting from the removal of a step point-torque. Again, the Fourier transform of the autocorrelation 
yields the noise spectrum. 
\begin{table}[tbp]
\begin{center}
\begin{tabular}
[c]{l@{\hspace{0.1cm}}l@{\hspace{0.1cm}}l@{\hspace{0.1cm}}l@{\hspace{0.1cm}}l@{\hspace{0.1cm}}
l@{\hspace{0.1cm}}l@{\hspace{0.1cm}}l@{\hspace{0.1cm}}l}
$ $ & $L (\mu$m) & $w (\mu$m) & $h (\mu$m) & $k$ (N/m) & $k_t$ (N-m/rad) & $f_0$ (kHz)
\\ \hline \hline \\
(1) & 197 & 29    & 2     & 1.3 & $1.6 \times 10^{-8}$   & 71 \\
(2) & 140 & 15.6 & 0.6 & 0.1 & $8.9 \times 10^{-10}$ & 38 \\
\end{tabular}
\end{center}
\caption{Summary of the cantilever geometries and material properties.  
(1) The rectangular cantilever. (2) The V-shaped cantilever used is the commercially 
available Veeco MLCT Type E microlever that is used in AFM~\cite{veeco}. The 
geometry is given by the cantilever length $L$, width $w$, and height $h$. 
For the V-shaped cantilever the total length between the two arms at the base 
is $b=161.64\mu$m. The cantilever spring constant $k$, torsional spring constant $k_t$, 
and resonant frequency in vacuum $f_0$ are determined using finite element numerical 
simulations.  The cantilevers are immersed in water with density 
$\rho_l$ = 997 kg/$\text{m}^3$ and dynamic viscosity 
$\eta$ = 8.59 $\times$ $10^{-4}$ kg/m-s.}
\label{table:geometry}
\end{table}

A powerful aspect of this approach is that it is possible to use deterministic numerical simulations to 
determine $U_{1}(t)$ and $\Theta_{1}(t)$ for the precise cantilever geometries and conditions of 
experiment. This includes the full three-dimensionality of the dynamics which are not accounted for 
in  available theoretical descriptions.  The numerical results can be used to guide the 
development of more accurate theoretical models. 

\section{The stochastic dynamics of cantilever tip-deflection and tip-angle}
\label{section:stochastic-dynamics}

The stochastic dynamics of the cantilever tip-displacement  $u_1(t)$ and that of the 
tip-angle $\theta_1(t)$ yield interesting differences. Using the thermodynamic approach, 
insight into these differences can be gained by performing a mode expansion of the 
cantilever using the initial deflection required by the deterministic calculation. The two 
cases of a tip-force and a tip-torque result in a significant difference in the mode 
expansion coefficients which can be directly related to the resulting stochastic dynamics.

For small deflections the dynamics of a cantilever with a non-varying cross section are given by the 
Euler-Bernoulli beam equation, 
\begin{equation}
\mu \frac{\partial^2 U}{\partial t^2} + E I \frac{\partial^4 U}{\partial x^4} = 0,
\end{equation}
where $U(x,t)$ is the transverse beam deflection, $\mu$ is the mass per unit length, $E$ is Young's modulus, 
and $I$ is the moment of inertia~\cite{landau:1959}. For the case of a cantilever where a step force has been applied to the 
tip at some time in the distant past the steady deflection of the cantilever at $t=0$ is given by
\begin{equation}
U(x) = - \frac{F_0}{2 E I} \left( \frac{x^3}{3} - L x^2 \right),
\label{eq:initial_def_force} 
\end{equation}
where $L$ is the length of the cantilever and the appropriate boundary conditions are 
$U(0)=U'(0)=U''(L)=0$ and $U'''(L) = - F_0/E I$.  The prime denotes differentiation with respect to $x$. 

Similarly, the deflection of the same cantilever beam due to the application of a point-torque at 
the cantilever-tip is quadratic in axial distance and is given by
\begin{equation}
U(x) = \frac{\tau_0}{2 E I} x^2,
\label{eq:initial_def_torque}
\end{equation}
where the appropriate boundary conditions are $U(0)=U'(0)=U'''(L) = 0$ and $U''(L)=\tau_0/E I$. The 
angle of the cantilever measured relative to the horizontal or undisplaced cantilever is then 
given by $\tan \Theta = U'(x)$.

The mode shapes for a cantilevered beam are given by
\begin{eqnarray}
\Phi_n(x) &=& -\left(\cos \kappa L + \cosh \kappa L\right) \left(\cos \kappa x - \cosh \kappa x\right)  \nonumber \\
                 &-& \left(\sin \kappa L - \sinh \kappa L\right) \left(\sin \kappa x - \sinh \kappa x\right),  
\end{eqnarray}
where $n$ is the mode number, and the characteristic frequencies are given by $\kappa^4=\omega^2 \mu/EI$.  The mode numbers $\kappa$  
are solutions to $1 + \cos{\kappa L} \cosh{\kappa L} = 0$~\cite{landau:1959}. The initial cantilever displacement given by 
Eqs.~(\ref{eq:initial_def_force}) and~(\ref{eq:initial_def_torque}) can be expanded into the beam modes
\begin{equation}
U(x) = \displaystyle\sum_{n=1}^{\infty} a_n \Phi_n(x), 
\end{equation}
with mode coefficients $a_n$.  The total energy $E_b$ of the deflected beam is given by 
\begin{equation}
E_{b} = \frac{E I}{2} \int_0^L U''(x)^2 dx,
\end{equation}
which is entirely composed of bending energy.  The fraction of the total bending energy contained 
in an individual mode is given by
\begin{equation}
b_n = \frac{E I}{2 E_b} \int_0^L \left( a_n \Phi_n''(x) \right)^2 dx. 
\end{equation}
The coefficients $b_n$ for the rectangular cantilever of Table~\ref{table:geometry} are shown 
in Table~\ref{table:energy}.  For the case of a force applied to the cantilever tip, 97\% of the total 
bending energy is contained in the fundamental mode and the energy contained in the higher modes 
decays rapidly with less than 1\% of the energy contained in mode three. When a point-torque is applied to 
the same beam it is clear that a significant portion of the bending energy is spread over the higher modes. 
Only 61\% of the energy is contained in the fundamental mode and the decay in energy with mode number 
is more gradual. The fifth mode for the tip-torque case contains more energy than the second mode for the 
tip-force case. Although we have only discussed a mode expansion for the rectangular cantilever, the 
V-shaped cantilever will exhibit similar trends since the transverse mode shapes are similar to that of a 
rectangular beam~\cite{stark:2001}.

The variation in the energy distribution among the modes required to describe the initial deflection of the 
cantilever can be immediately connected to the resulting stochastic dynamics.  For the deterministic calculations 
the initial displacement can be arbitrarily set to a small value. In this limit the modes of the cantilever beam are not 
coupled through the fluid dynamics. As a result, the stochastic dynamics of each mode can be treated as the ring 
down of that mode from the initial deflection. This indicates that the more energy that is distributed amongst the 
higher modes initially the more significant the ring down and, using the fluctuation-dissipation theorem, the more 
significant the stochastic dynamics. 

The mode expansion clearly shows that the tip-torque case has more energy 
in the higher modes. This suggests that stochastic measurements of 
the cantilever tip-angle will have a stronger 
signature from the higher modes than measurements of cantilever tip-displacements.  
Using finite element simulations for the precise geometries of interest we 
quantitatively explore these predictions.
\begin{table}[tbp]
\begin{center}
\begin{tabular}
[c]{l@{\hspace{0.5cm}}l@{\hspace{0.5cm}}l@{\hspace{0.5cm}}l}
$n$ & $b_n$ (tip-force)  & $b_n$ (tip-torque)
\\ \hline \hline \\
$1$ & 0.97068    & 0.61308 \\
$2$ & 0.02472    & 0.18830 \\
$3$ & 0.00315    & 0.06473 \\
$4$ & 0.00082    & 0.03309 \\
$5$ & 0.00030    & 0.02669 \\
\end{tabular}
\end{center}
\caption{The fraction of the total energy $E_b$ contained in the first five beam modes 
given by the coefficients $b_n$. The tip-force results are for a rectangular beam that 
has been deflected by the application of a point force to the cantilever tip.  The 
tip-torque results are for a rectangular beam that has been deflected 
by the application of a point torque to the cantilever tip. The coefficients clearly show 
that the tip-torque case has significantly more energy contained in the higher modes.}
\label{table:energy}
\end{table}
\begin{table}[tbp]
\begin{center}
\begin{tabular}
[c]{l@{\hspace{0.4cm}}l@{\hspace{0.2cm}}l@{\hspace{0.3cm}}l}
$ $ & $\left< u_{1}^2 \right>^{1/2}$(nm) & $\left< \theta_{1}^2 \right>^{1/2}$(rad)
\\ \hline \hline \\
(1) & $5.6$ & $5.0 \times 10^{-7}$ \\
(2) & $20$ & $7.0 \times 10^{-9}$  \\
\end{tabular}
\end{center}
\caption{The magnitude of stochastic fluctuations in tip-deflection and in 
tip-angle for the rectangular~(1) and V-shaped~(2) cantilevers.  These values were 
obtained from numerical simulations simulations of the beams in vacuum.}
\label{table:stchmag}
\end{table}

\section{The stochastic dynamics of a rectangular cantilever}
\label{section:rectangular-cantilever}

We have performed deterministic numerical simulations of the three-dimensional, time dependent, 
fluid-solid interaction problem to quantify the stochastic dynamics of a rectangular  
cantilever immersed in water using the thermodynamic approach discussed in 
Section~\ref{section:thermo}. The deterministic numerical simulations are done using a 
finite element approach that is described elsewhere~\cite{yang:1994,ESI}.

The stochastic fluctuations in cantilever tip-displacement for
a rectangular cantilever in water have been described elsewhere
~\cite{chon:1999,paul:2004,paul:2006,clark:2006}. 
In the following we compare these results with the stochastic dynamics as determined 
by the fluctuations of the cantilever tip-angle. The geometry of the the specific micron scale 
cantilever we explore is given in Table~\ref{table:geometry}.

\begin{figure}[htb]
  \begin{center}
    \includegraphics[width=3.0in]{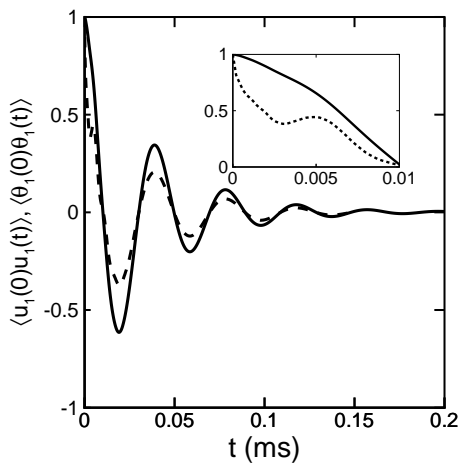}
  \end{center}
  \caption{The normalized autocorrelation of the rectangular cantilever
  for tip-deflection (solid) and tip-angle (dashed).  (Inset) 
  A detailed view of the autocorrelation at short time differences to
  illustrate the influence of higher modes in the tip-angle measurements.}
  \label{fig:rectangular-auto}
\end{figure}

As discussed in Section~\ref{section:thermo} the autocorrelations in equilibrium fluctuations 
follow immediately from the ring down of the cantilever due to the removal of a step force 
(to yield $\left< u_1(0)u_1(t) \right>$) or step point-torque (to yield $\left< \theta_1(0) \theta_1(t) \right>$). 
The autocorrelations of the rectangular cantilever 
are shown in Fig.~\ref{fig:rectangular-auto}. The magnitude of the noise 
is quantified by the root mean squared tip-angle and deflection which is listed in 
Table~\ref{table:stchmag}. 

A comparison of the autocorrelations yields some interesting 
features. At short times $\left< \theta_1(0) \theta_1(t) \right>$ shows the presence of higher harmonic 
contributions. This is shown more clearly in the inset of Fig.~\ref{fig:rectangular-auto}. This 
further suggests that the angle autocorrelations are more sensitive to higher mode dynamics as discussed in 
Section~\ref{section:stochastic-dynamics}.

\begin{figure}[htb]
  \begin{center}
    \includegraphics[width=2.75in]{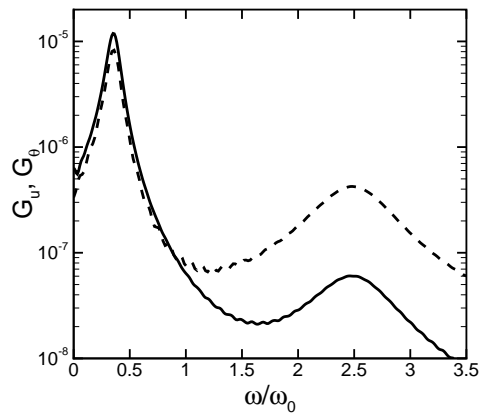}
  \end{center}
  \caption{The noise spectra of stochastic fluctuations in cantilever
  tip-angle (dashed) and tip-deflection (solid) for the rectangular cantilever. 
  The curves are normalized to have the same area, however only the first two modes 
  are shown.}
  \label{fig:rectangular-noise}
\end{figure}

The Fourier transform of the autocorrelations yield the noise spectra shown in 
Fig.~\ref{fig:rectangular-noise}. In our notation the subscript of $G$ indicates the 
variable over which the noise spectrum is measured: $G_\theta$ is the noise spectrum 
for tip-angle and $G_u$ is the noise spectrum for tip-displacement.  The equipartition 
theorem of energy yields,
\begin{eqnarray}
 \frac{1}{2 \pi} \int^{\infty}_{0} G_u (\omega) d\omega &=& \frac{k_B T}{k} \\ 
  \frac{1}{2 \pi} \int^{\infty}_{0} G_\theta  (\omega) d\omega &=& \frac{k_B T}{k_t}
 \label{eq:equipartition}
\end{eqnarray}
where $k$ and $k_t$ are the transverse and torsional spring constants,
respectively.  The curves in Fig.~\ref{fig:rectangular-noise} are normalized using the
equipartition result to 
have a total area of unity. Using this normalization the area under a peak is an 
indication of the amount of energy contained in a particular mode. 
Figure~\ref{fig:rectangular-noise} shows only the first two modes, 
although the numerical simulations include all of the modes (within the numerical 
resolution of the finite element simulation). The energy distribution across the 
first two modes shows the significance of the second mode for the tip-angle dynamics.

Using a simple harmonic oscillator approximation it is straight forward to compute the peak 
frequency $\omega_f$ and quality $Q$ for the cantilever in fluid. Using a single mode approximation 
yields the values shown in Table~\ref{table:rectangular-noise}. As expected there is a significant 
reduction in the cantilever frequency when compared with the resonant frequency in 
vacuum $\omega_0$ and the quality factor is quite low because of the strong fluid dissipation. 
The values of $\omega_f$ and $Q$ for tip-angle and tip-dispacement are nearly equal.  This 
is expected since the displacements and angles are very small, resulting in negligible coupling 
between the modes. Any differences in $\omega_f$ and $Q$ can be attributed to using a 
single mode approximation. 

It is useful to compare these results with the commonly used approximation of an oscillating,
infinitely long cylinder with radius $w/2$~\cite{sader:1998,tuck:1969,paul:2006}. The cantilever 
used here has an aspect ratio of $L/w \approx 7$ and the infinite cylinder theory is quite good at 
predicting of $\omega_f$ and $Q$.
\begin{table}[h]
\begin{center}
\begin{tabular} [c]{l@{\hspace{1cm}}l@{\hspace{1cm}}l@{\hspace{1cm}}}
$ $ & $\omega_f / \omega_0$ &  $Q$  \\ \hline \hline \\
(1) & 0.35 & 3.34 \\
(2) & 0.36 &  3.26 \\
\end{tabular}
\end{center}
\caption{The peak frequency and quality factor of the fundamental mode 
of the rectangular cantilever determined by finite element simulations using the 
thermodynamic approach. (1) is computed using the cantilever tip-displacement 
due to the removal of a step force. (2) is computed using the cantilever tip-angle 
due to the removal of a point-torque. The frequency result is normalized by the resonant 
frequency in vacuum $\omega_0$. Using the infinite cylinder approximation with 
a radius of $w/2$ the analytical predictions are $Q=3.24$ and $\omega_f/\omega_0=0.34$. }
\label{table:rectangular-noise}
\end{table}

\section{The stochastic dynamics of a V-shaped cantilever}

We now explore the stochastic dynamics of a V-shaped cantilever in fluid. An integral 
component of any theoretical model is an analytical description of the resulting 
fluid flow field caused by the oscillating cantilever. The deterministic finite element 
simulations that we performed yield a quantitative picture of the resulting fluid dynamics. 
Exploring the flow fields further yields insight into the dominant features that contribute 
to the cantilever dynamics.
\begin{figure}[htb]
  \begin{center}
    \includegraphics[width=3.0in]{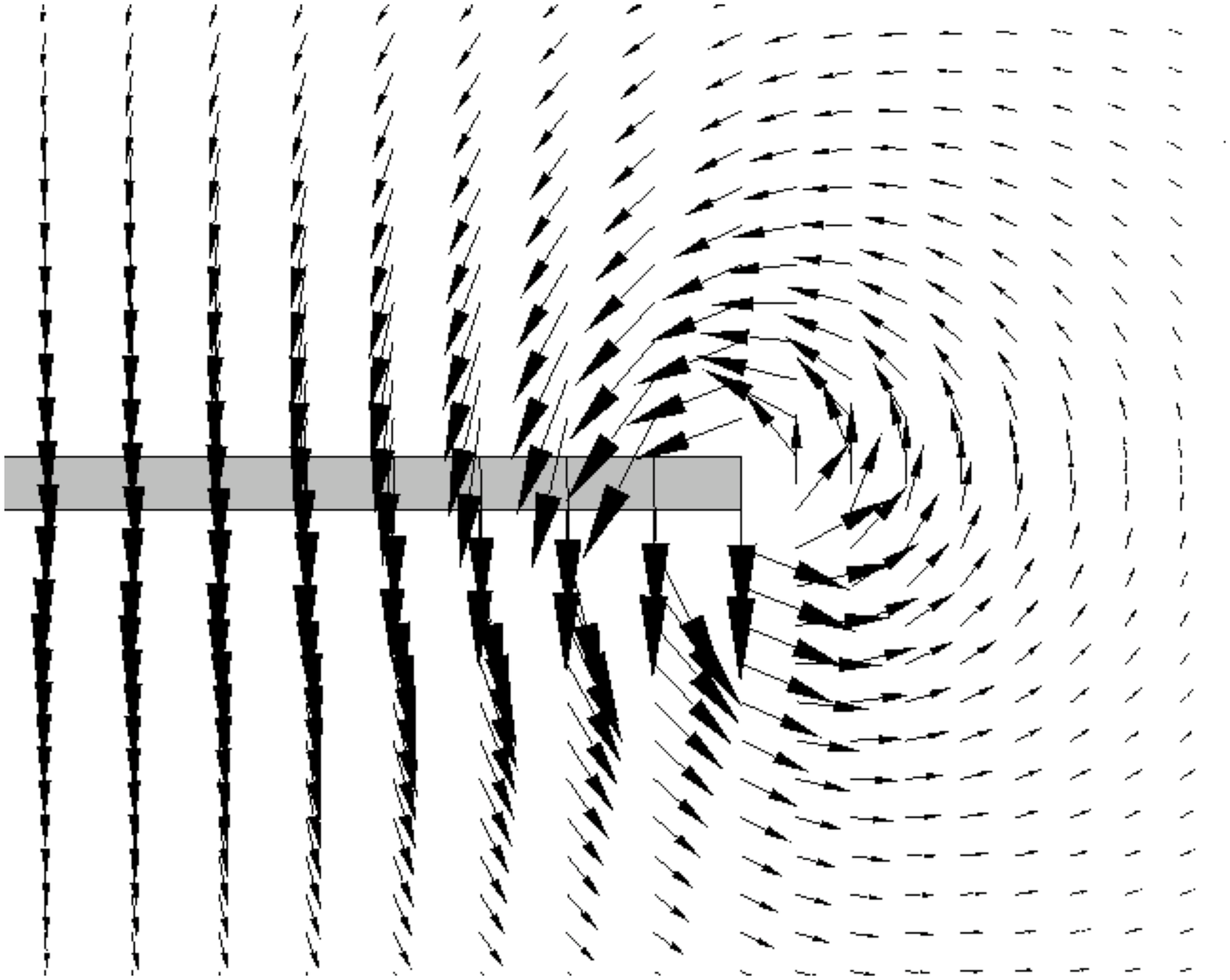} \\
    \vspace{0.2cm}
    \includegraphics[width=3.0in]{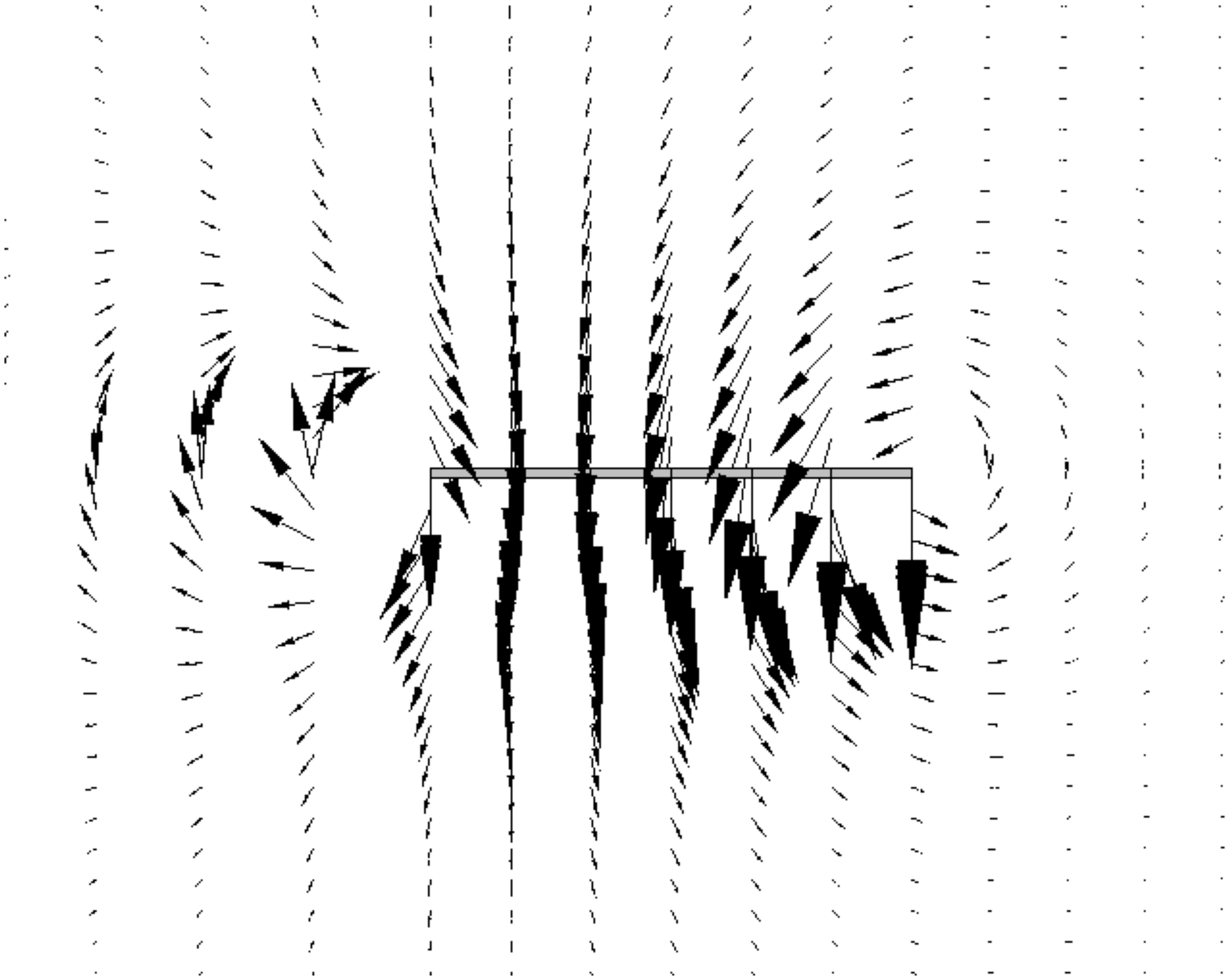}
  \end{center}
  \caption{The fluid flow near the tip of the cantilever as illustrated by the velocity 
  vector field calculated from finite element numerical simulations. A cross section 
  of the $x-y$ plane at $z=0$ is shown (see Fig.~\ref{fig:beam}) that is a close-up 
  view of the tip-region. The shaded region indicates the cantilever (because of the 
  small deflections used in the simulations that cantilever does not appear to be 
  deflected). (top) The flow field near the tip of the rectangular cantilever.  This flow
  field is at t=6$\mu$s and the magnitude of the largest velocity vector shown
  is -0.3 nm/s. 
  (bottom) The flow field near the tip of the V-shaped cantilever.  This flow field 
  is at t=7.2$\mu$s and the magnitude of the largest velocity vector shown is -26 nm/s.
 The shaded region 
  indicates the tip region where the two single arms have merged. The open region 
  to the left is where the two single arms have separated revealing the open region 
  in the interior of the V-shaped cantilever.}
  \label{fig:tip-flow}
\end{figure}

As discussed earlier, for long and slender rectangular cantilevers the flow field is 
often approximated by that of a cylinder of diameter $w$ undergoing 
transverse oscillations. This approach assumes that the fluid flow is essentially 
two-dimensional in the $y-z$ plane and neglects any flow over the tip of the 
cantilever. Figure~\ref{fig:tip-flow} (top) illustrates this tip flow for the rectangular
cantilever using vectors of
the fluid velocity in the $x-y$ plane at $z=0$. The figure is a close-up view near 
the tip of the cantilever. It is evident that the flow over the rectangular cantilever 
is nearly uniform in the axial direction leading up to the tip. However, near the 
tip there is a significant tip flow that decays rapidly in the axial direction away 
from the tip. The increasing significance of the tip flow as 
the cantilever geometry becomes shorter (for example, by simply decreasing $L$) 
is not certain and remains an interesting open question. However, for the geometry 
used here it is clear that this tip-flow is negligible based upon the accuracy 
of the analytical predictions using the two-dimensional model.

Figure~\ref{fig:tip-flow} (bottom) illustrates the tip flow for the V-shaped 
cantilever, again by showing velocity vectors in the $x-y$ plane at $z=0$. The 
shaded region indicates the part of the cantilever where the two
arms have merged. To the right of the shaded region illustrates flow off the tip and to the 
left indicates flow that circulates back in between the two individual arms.

\begin{figure}[htb]
  \begin{center}
    \includegraphics[width=3.0in]{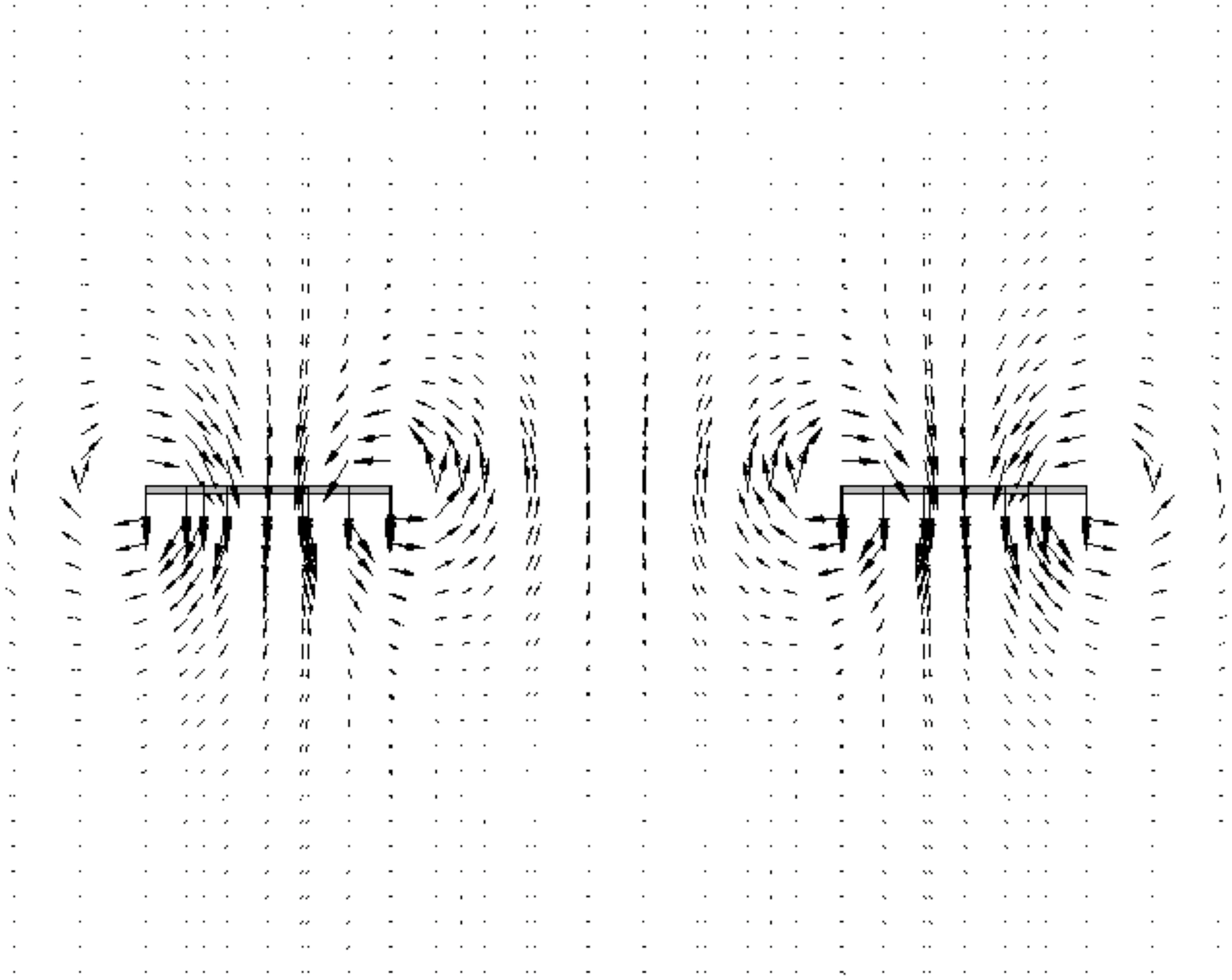} \\
    \vspace{0.2cm}
    \includegraphics[width=3.0in]{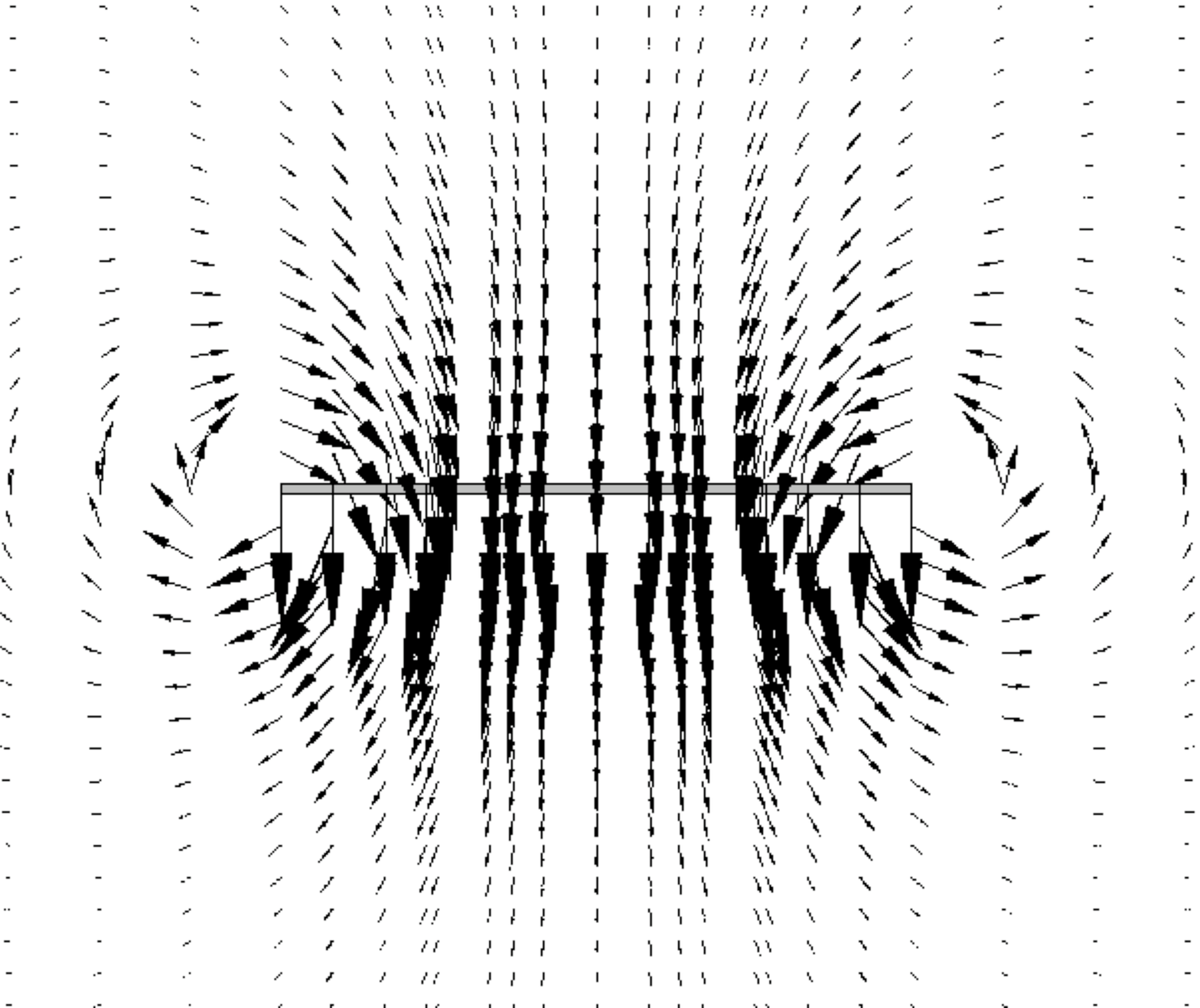}
  \end{center}
  \caption{The fluid velocity vector field at two axial positions along the 
  V-shaped cantilever calculated from deterministic finite element numerical simulations. 
  Cross sections  of the $y-z$ plane are shown (see Fig.~\ref{fig:beam}), the 
  entire simulation domain is not shown and the shaded region indicates the 
  cantilever. Both images are taken at t=7.2$\mu$s and the
  maximum velocity vector shown is -26 nm/s.  (top) The $y-z$ plane 
  at $x=77\mu$m. The skewed width of a single arm 
  of the cantilever in this cross-section is $18\mu$m. The distance separating the 
  two cantilever arms is $36\mu$m. (bottom) The $y-z$ plane at $x=108.8\mu$m. 
  This is the point at which the two single arms join to make a continuous cross-section 
  of width $2w$.}
  \label{fig:cross-flow}
\end{figure}

In order to illustrate the three-dimensional nature of this flow, the flow field 
in the $y-z$ plane is shown at two axial locations in Fig.~\ref{fig:cross-flow}. 
Figure~\ref{fig:cross-flow}(top) is at axial location $x=77\mu$m. The two shaded 
regions indicate the two arms of the cantilever. Each arm is generating a flow 
with a viscous boundary layer (Stokes layer) as expected from previous 
work on rectangular cantilevers. However, the Stokes  layers interact in a 
complicated manner near the center. It is expected 
that as one goes from the base of the cantilever to the tip that these fluid 
structures would transition from non-interacting to strongly-interacting.

Figure~\ref{fig:cross-flow}(bottom) illustrates the flow field at axial location 
$x=108.8\mu$m, the axial location at which the two arms of the 
cantilever merge to form the tip region. The length of the shaded region 
is therefore $36\mu$m or twice that of a single arm shown in 
Fig.~\ref{fig:cross-flow}(top). For this tip region the flow field is similar to 
what would be expected of a single rectangular cantilever of this width. 

Overall, it is clear that the fluid flow field is more complex for the 
V-shaped cantilever than for the long and slender rectangular 
beam. For the V-shaped cantilever the flow is three-dimensional near the tip 
region where the two arms join together.

Central to the flow field dynamics are the interactions of the two Stokes layers  
caused by the oscillating cantilever arms. The thickness of these Stokes layers are expected 
to scale with the frequency of oscillation as $\delta_s/a \sim {R_\omega}^{-1/2}$ where $a$ is the half-width of 
the cantilever and $R_\omega = \omega a^2/\nu$ is a frequency based Reynolds number (often 
called the frequency parameter).    For the relevant case of a cylinder of radius $a$ oscillating at 
frequency $\omega$ 
the solution to the unsteady Stokes equations yields a distance
of approximately $5 \delta_s$ 
to capture 99\% of the fluid velocity in the viscous boundary layer~\cite{carvajal:2007}. For 
a single arm of the V-shaped cantilever this distance is nearly $10\mu$m. 
In comparison, the total distance between the two arms at the base is $125\mu$m. This 
separation is large enough such that the two Stokes layers have negligible interactions 
near the base. However, as the arms approach one another with axial distance the 
Stokes layers overlap and eventually merge at the tip.

\begin{figure}[htb]
  \begin{center}
    \includegraphics[width=3.0in]{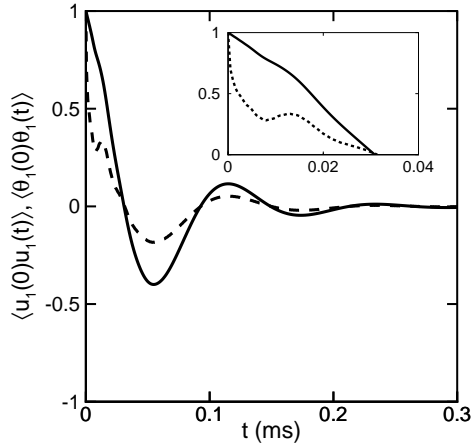}
  \end{center}
  \caption{The normalized autocorrelation of equilibrium fluctuations in the tip-deflection 
  $\left<u_1(0)u_1(t)\right>$ (solid lined) and in tip-angle 
  $\left<\theta_1(0)\theta_1(t)\right>$ (dashed-line) 
  for the V-shaped cantilever. The inset shows a close-up of the dynamics for short 
  time differences to illustrate the influence of the higher modes in the tip-angle measurements.}
  \label{fig:vshape-auto}
\end{figure}

\begin{figure}[htb]
  \begin{center}
    \includegraphics[width=2.75in]{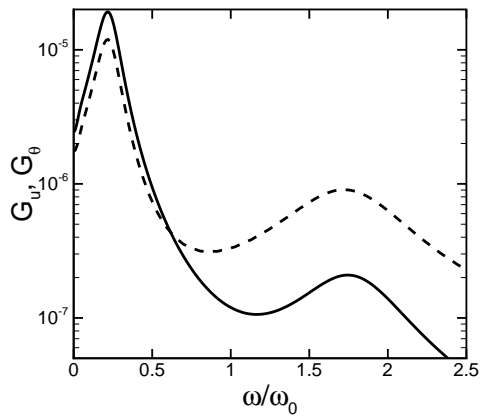}
  \end{center}
  \caption{The noise spectra for the V-shaped cantilever as determined from the 
  tip-displacement $G_x$ (solid line) and from tip-angle $G_\theta$ (dashed line).
  The curves are normalized to have an area of unity, with only the first
  two modes shown.}
  \label{fig:vshape-noise}
\end{figure}

Despite the complicated interactions of the three-dimensional flow caused by the cantilever tip and 
the axial merging of the two Stokes layers, the V-shaped cantilever behaves as a damped 
simple harmonic oscillator. The autocorrelations in tip-angle and tip-displacement 
that are found using full finite element numerical simulations are shown
in Fig.~\ref{fig:vshape-auto}. It is again 
clear that the tip-angle dynamics have significant contributions from the higher modes, see the 
inset of Fig.~\ref{fig:vshape-auto}.  The area normalized noise spectra are shown 
in Fig.~\ref{fig:vshape-noise}.

Using a simple harmonic oscillator analogy a peak frequency  
and a quality factor can be determined from the first mode in the noise spectra of 
Fig.~\ref{fig:vshape-noise}.  These values are given in the first two rows of 
Table~\ref{table:noise-vshape}. The quality of the cantilever is $Q\approx2$
and the peak frequency is reduced significantly, $\omega_f/\omega_0\approx0.2$.
compared to the resonant frequency 
in the absence of a surrounding viscous fluid.

It is insightful and of practical use to determine the geometry of the equivalent rectangular 
beam that would yield the precise values of $k$, $\omega_f$, and $Q$ 
calculated for the V-shaped cantilever from full 
finite-element numerical simulations. For the rectangular beam the 
equations are well known (c.f. Ref.~\cite{paul:2006}) and yield a unique value of length 
$L'$, width $w'$, and height $h'$ as shown below,
\begin{eqnarray}
k &=& \frac{3 E I}{L'^3} = \frac{E w' h'^3}{4 L'^3}, \label{eq:k} \\ 
Q &=& \frac{m_f \omega_f}{\gamma_f} = \frac{\frac{4 \rho_c h'}{\pi \rho_f w'} + \Gamma'(w',\omega_f)}{\Gamma''(w',\omega_f)}, \label{eq:Q}
\end{eqnarray}
where the peak frequency is determined from the maximum of the noise spectrum,
\begin{eqnarray}
G_u &=& \frac{4 k_B T}{k} \frac{1}{\omega_0} \label{eq:omega} \\ &\times&\frac{T_0 \tilde{\omega} \Gamma''(R_0 \tilde{\omega})}{\left[ (1-\tilde{\omega}^2 (1 + T_0 \Gamma'(R_0 \tilde{\omega})))^2 + (\tilde{\omega}^2 T_0 \Gamma''(R_0 \tilde{\omega}))^2 \right] }. \nonumber
\end{eqnarray}
In the above equations $\tilde{\omega} = \omega/\omega_0$ is the normalized frequency,
$\alpha=0.234$ is a constant factor to determine an equivalent lumped mass 
for a rectangular beam, $m_f$ is the equivalent mass of the cantilever plus the 
added fluid mass, $\gamma_f$ is the fluid damping, $\Gamma$ is the hydrodynamic 
function for an infinite cylinder, $\Gamma'$ is the real part of $\Gamma$, and $\Gamma''$ is 
the imaginary part of $\Gamma$. Equations~(\ref{eq:k})-(\ref{eq:Q}) can be solved to 
yield values for the unknown geometry of the equivalent 
rectangular beam $L'$, $w'$, and $h'$ which are given in Table~\ref{table:vshape-virtual-beam}. 
The equivalent beam is shorter, thinner, and wider than the V-shaped cantilever. Importantly, 
the width of the equivalent beam is nearly twice that of a single arm of the V-shaped 
cantilever.
\begin{table}[h]
\begin{center}
\begin{tabular} [c]{l@{\hspace{1cm}}l@{\hspace{1cm}}l@{\hspace{1cm}}}
$L'/L$ & $w'/w$ &  $h/h'$  \\ \hline \hline \\
 0.8 & 1.9 & 0.8 \\
\end{tabular}
\end{center}
\caption{The geometry of the equivalent rectangular beam that yields the exact 
values of $k$, $\omega_f$, and $Q$ for the V-shaped cantilever that have been 
determined from full finite-element numerical simulations. The length, width, and height 
of the equivalent beam $(L',w',h')$ are calculated using Eqs.~(\ref{eq:k})-(\ref{eq:Q}) 
and are normalized by the values of ($L,w,h$) for the V-shaped cantilever given 
in Table~\ref{table:geometry}.}
\label{table:vshape-virtual-beam}
\end{table}

These results suggest that the parallel beam approximation 
(PBA)~\cite{albrecht:1990,sader:1995,sader:1993,butt:1993} commonly used to determine 
the spring constant for a V-shaped cantilever may also provide a useful geometry for determining 
the dynamics of V-shaped cantilevers in fluid. In this approximation the V-shaped cantilever 
is replaced by an equivalent rectangular beam of length $L$, width $2w$, and height $h$ 
to yield a simple analytical expression for the spring constant. This has been shown to 
be quite successful for V-shaped cantilevers that have arms that are not significantly 
skewed. The results of using the geometry of this approximation to determine $\omega_f$ 
and $Q$ from the two-dimensional cylinder approximation are 
shown on the third row of Table~\ref{table:noise-vshape}. It is clear that this is 
quite accurate. It is expected that these results will remain 
useful for cantilever geometries that do not deviate significantly from that of an equilateral 
triangle as studied here. An exploration of the breakdown of this approximation is possible using the 
methods described but is beyond the scope of the current efforts.
\begin{table}[h]
\begin{center}
\begin{tabular} [c]{l@{\hspace{1cm}}l@{\hspace{1cm}}l@{\hspace{1cm}}}
$ $ & $\omega_f / \omega_0$ &  $Q$  \\ \hline \hline \\
(1)           & 0.21 &  1.98 \\
(2)           & 0.22 &  2.04 \\
(L,2w,h) & 0.19 &  1.98 \\
\end{tabular}
\end{center}
\caption{The peak frequency and quality factor of the fundamental mode 
of the V-shaped cantilever determined by finite element simulations using the 
thermodynamic approach. (1) is computed using the cantilever tip-displacement 
due to the removal of a step force. (2) is computed using the cantilever tip-angle 
due to the removal of a point-torque. The third line represents theoretical predictions using 
the geometry of an equivalent rectangular beam 
given by $(L,2w,h)$. The frequency result is normalized by the resonant 
frequency in vacuum $\omega_0$.}
\label{table:noise-vshape}
\end{table}

\section{Quantifying the increased dissipation due to a planar boundary}

\begin{figure}[htb]
  \begin{center}
    \includegraphics[height=1.5in]{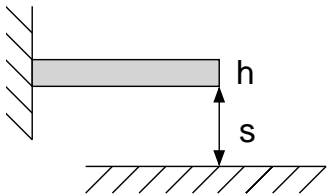}
  \end{center}
  \caption{A schematic of a cantilever a distance $s$ away
  from a solid planar surface (not drawn to scale). The cantilever undergoes flexural 
  oscillations perpendicular to the surface.}
  \label{fig:setup}
\end{figure}

In practice, the cantilever is never placed in an unbounded fluid and the 
influence of nearby boundaries must be accounted for to provide a complete 
description of the dynamics. In many cases the cantilever is purposefully brought 
near a surface out of experimental interest in order to probe some interaction with 
the cantilever or to probe the surface itself. To specify our discussion we will consider 
the situation depicted in Fig.~\ref{fig:setup} showing a cantilever 
a distance $s$ from a planar boundary. In the following we study the case 
where the cantilever exhibits flexural oscillations in the direction perpendicular 
to the boundary. However, we would like to emphasize that our approach 
is general and can be used to explore arbitrary cantilever orientations and oscillation 
directions if desired. The fluid is assumed 
to be unbounded in all other directions. It is well known that 
the presence of the boundary will influence the dynamics of the cantilever
~\cite{benmouna:2002,nnebe:2004,harrison:2007}. The 
result is a reduction in the resonant frequency and quality factor. This has been 
described theoretically for the case of a long and thin cantilever of simple 
geometry where the fluid dynamics have been assumed two-dimensional
~\cite{clarke:2005,clarke:2006,green:2005,green:2005:jap}.

\begin{figure}[htb]
  \begin{center}
    \includegraphics[width=2.75in]{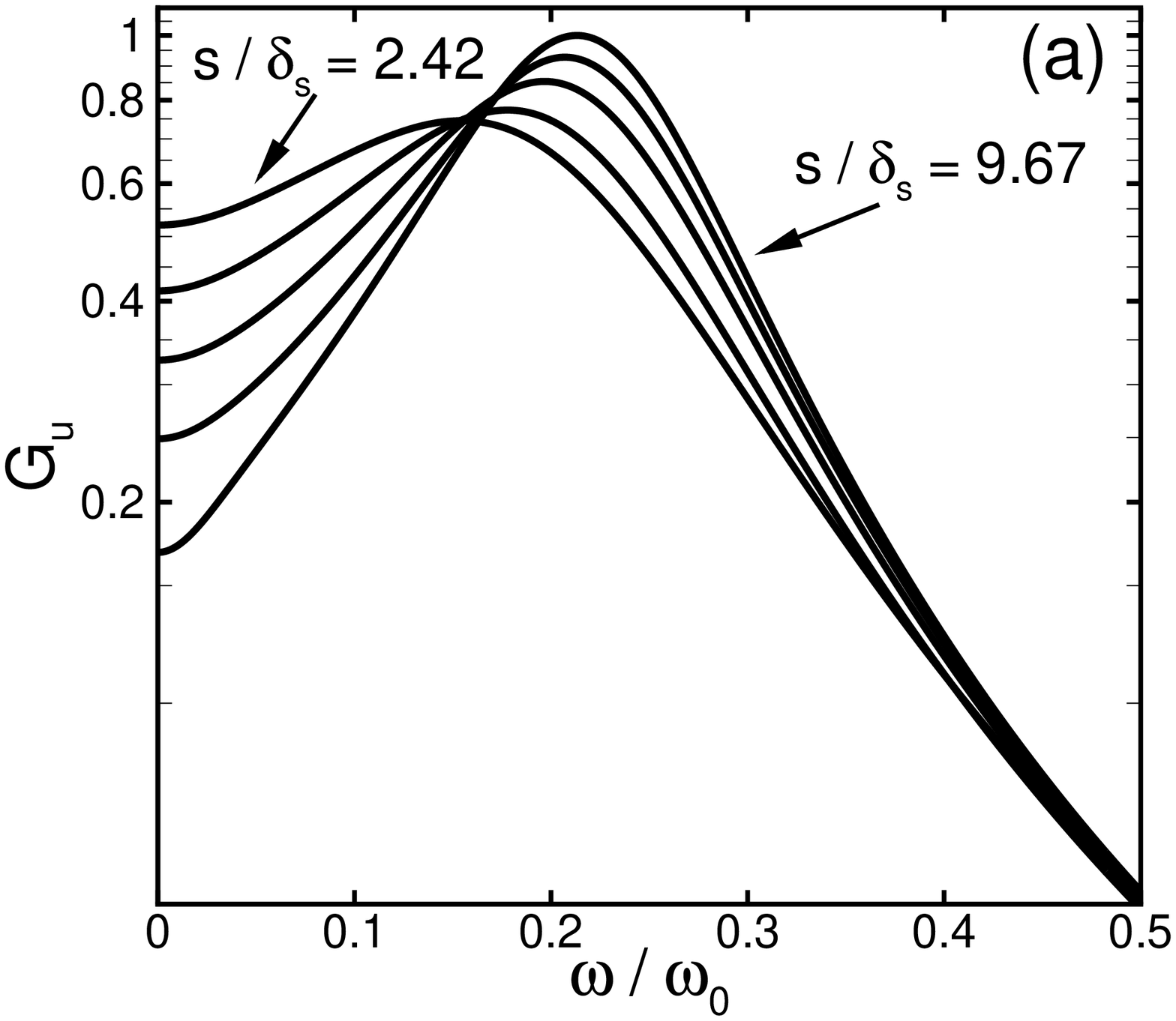} \\
    \includegraphics[width=2.75in]{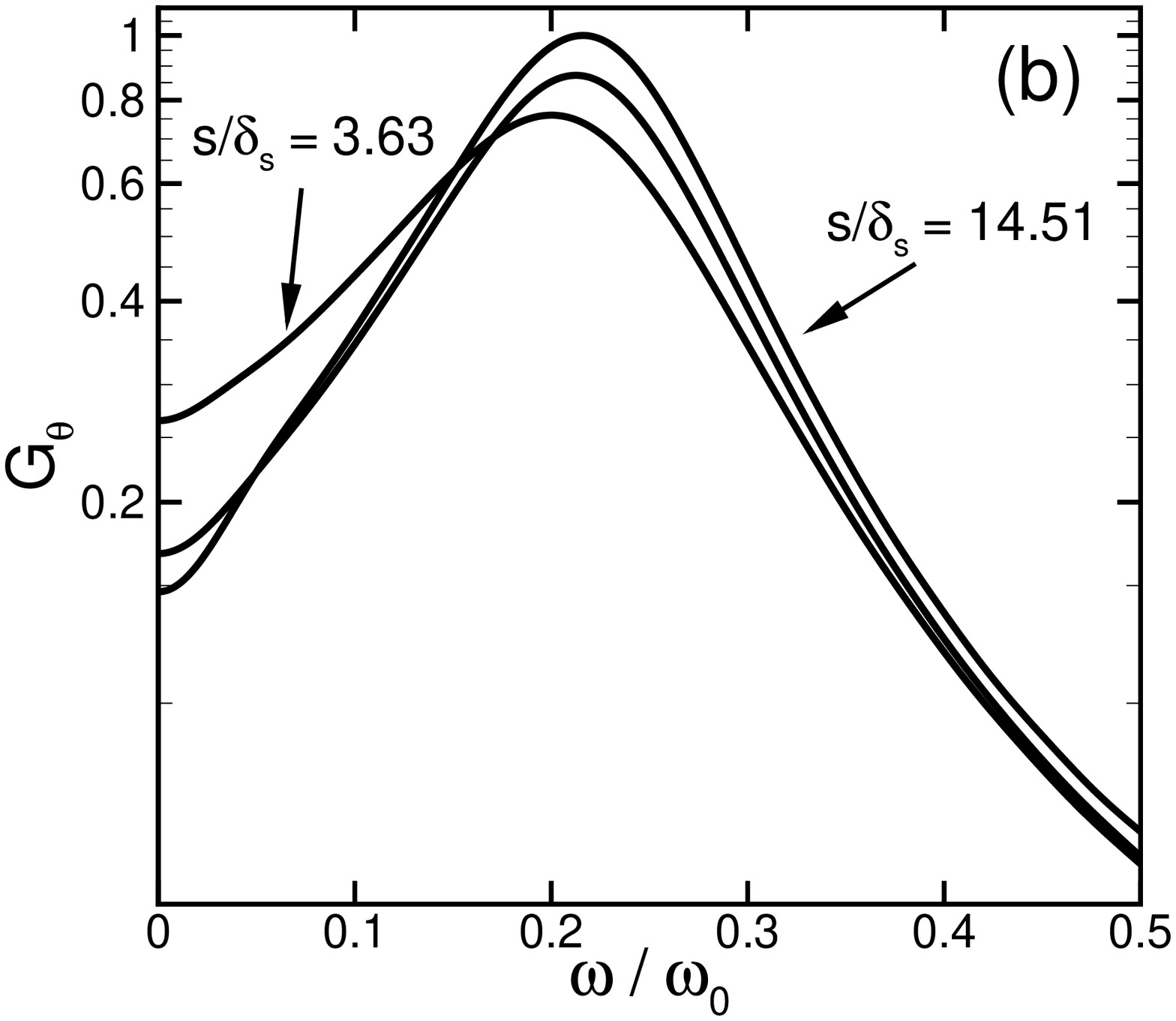}
  \end{center}
  \caption{Panel~(a) The noise spectra $G_u$ of stochastic fluctuations in cantilever
  tip-deflection for separations $s = 10,12,15,20,40\mu$m. Panel~(b) the noise 
  spectra $G_\theta$ of stochastic fluctuations in cantilever tip-angle for separations 
  $s = 15,25,60\mu$m. The spectra have been normalized by the maximum value of 
  $G_u$ or $G_\theta$. The smallest and largest values of separation are labeled 
  with all other values appearing sequentially.}
  \label{fig:psdsep}
\end{figure}

In the following we use the thermodynamic approach with finite element 
numerical simulations to quantify the dynamics of the V-shaped cantilever 
as a function of its separation from a boundary. We have performed 8 simulations 
over a range of separations from 10 to 60$\mu$m using both the tip-deflection 
and tip-angle formulations. The noise spectra for these simulations are shown 
in Fig.~\ref{fig:psdsep}. Using the insights from our simulations of the V-shaped 
cantilever in an unbounded fluid we expect the relevant length scale 
for the fluid dynamics to be twice the width of a single arm, $2w$. Using
the peak frequency of the V-shaped cantilever in unbounded
fluid yields a Stokes length $\delta_{s} = 4.14 \mu$m.  Scaling the separation
by the Stokes length yields. 
$2.5 \lesssim s/\delta_{s} \lesssim 15$ which covers the range from what is 
expected to be a strong influence of the wall to a negligible influence. 
Figure~\ref{fig:psdsep} clearly shows a reduction in the peak frequency and 
a broadening of the peak as the cantilever is brought closer to the boundary. 
In fact, for the smaller separations the peak is quite broad and the trend 
suggests that eventually the peak will become annihilated as the cantilever 
is brought closer to the boundary.

\begin{figure}[htb]
  \begin{center}
  \includegraphics[width=3.0in]{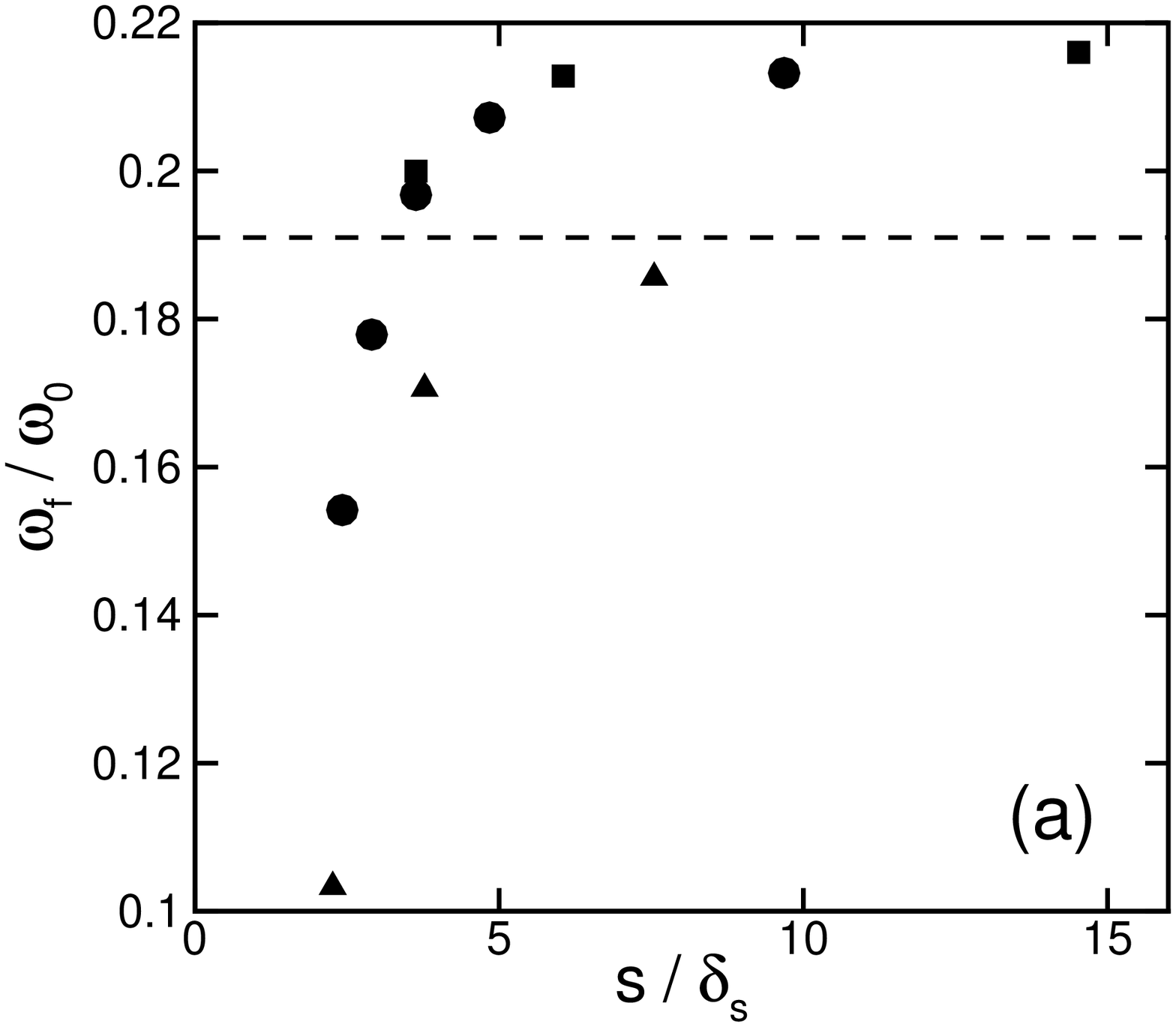} \\ 
   \includegraphics[width=3.0in]{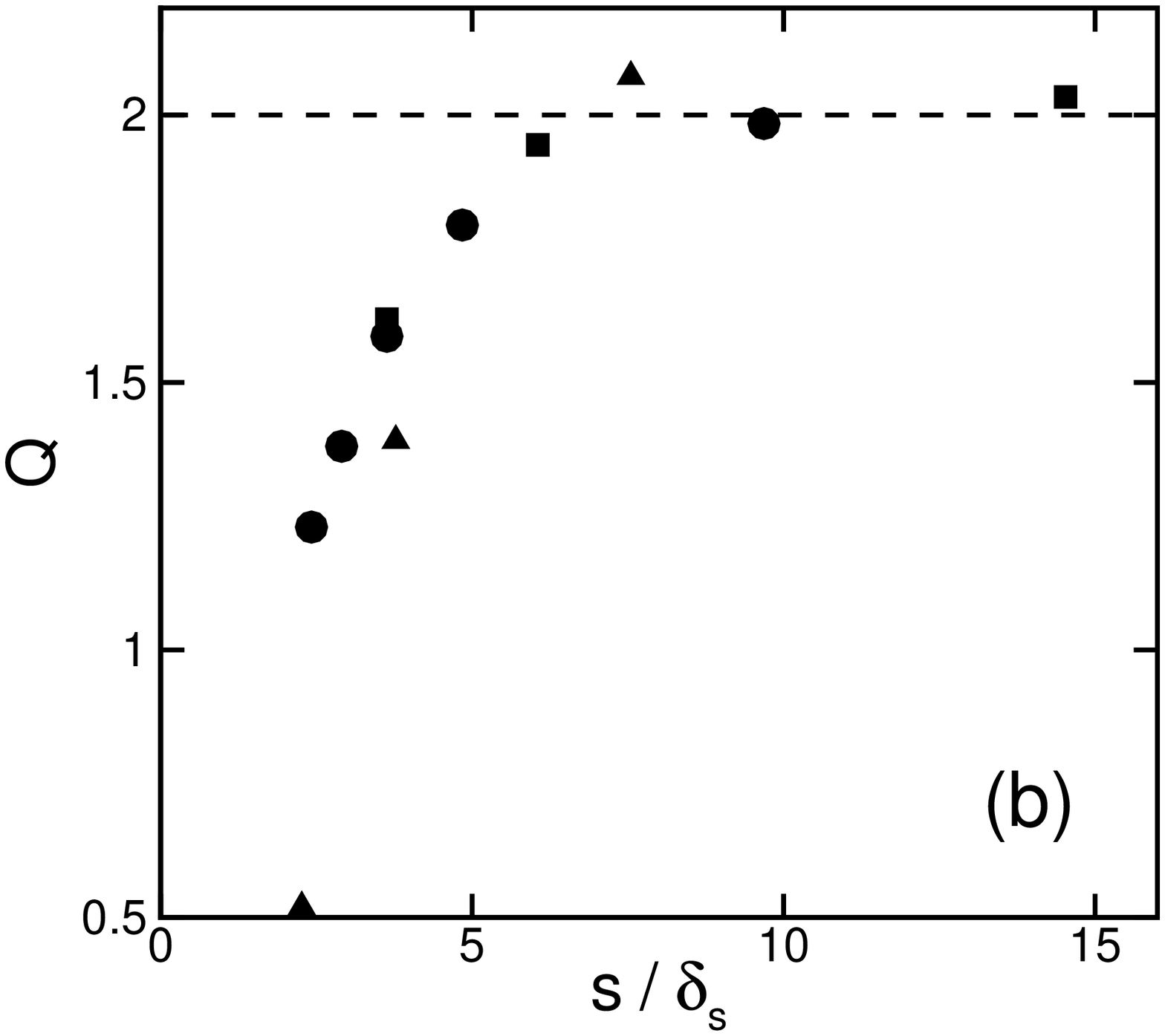}
  \end{center}
  \caption{The variation of the peak frequency (panel~(a)) and quality (panel~(b)) 
  of the fundamental mode of the V-shaped cantilever in fluid as a function of separation 
  from a nearby wall. Results calculated using tip-deflection are circles, results using 
  tip-angle are squares, and theoretical predictions using the results of 
  Ref.~\cite{green:2005:jap} are triangles. The peak frequency and quality factor of the 
  fundamental mode in an unbounded fluid are $\omega/\omega_f \approx 0.19$ 
  and $Q\approx2$ and are represented by the horizontal dashed line. The 
  distance $s$ is normalized by the Stokes length $\delta_s$ where 
  $a=w$ to yield $\delta_s=4.14\mu$m.}
  \label{fig:omfsep}
\end{figure}

Using the noise spectra we compute a peak frequency and 
a quality factor for the fundamental mode as a function of separation from the boundary, 
which are plotted in Fig.~\ref{fig:omfsep}. The horizontal dashed line represents the value of the 
peak frequency and quality factor in the absence of bounding surfaces using the 
two-dimensional infinite cylinder approximation~\cite{paul:2006} where the cylinder width has been 
chosen to be $2w$. It is clear from the results that for separations greater than 
$s/\delta_s \gtrsim 7$ the V-shaped cantilever is not significantly affected by the 
presence of the boundary. However, as the separation decreases below this value 
the peak frequency and quality factor decrease rapidly.

The triangles in Fig.~\ref{fig:omfsep} represent the theoretical predictions of Green and 
Sader~\cite{green:2005,green:2005:jap} using a two-dimensional approximation 
for a beam of uniform cross-section that accounts for the presence of the boundary. 
We have used a width of $2w$ in computing these theoretical predictions
for comparison with our numerical results. Despite 
the complex and three-dimensional nature of the flow field the theory is able to accurately 
predict the quality factor over the range of separations explored. The 
frequency of the peak for the V-shaped cantilever shows some deviation from these 
predictions.  

In general, an increase in the period of oscillation for a submerged
object can be attributed to the mass of fluid entrained by the 
object~\cite{stokes:1851}.  The lower peak frequency calculated for
the V-shaped cantilever using a two-dimensional solution indicates an
over-prediction of the mass loading.  This can be attributed to the 
three-dimensional flow around the tip being neglected for this approach.
It is reasonable to expect the cantilever tip to carry a smaller amount of 
fluid than a section of the beam body moving with the same velocity, 
see Fig.~\ref{fig:tip-flow}.  The quality factor relates to the ratio of the mass
loading and the viscous dissipation and is less sensitive to deviations
incurred from the two-dimensional approximation.  Despite neglecting
three-dimensional flow around the cantilever tip, the 
two-dimensional model for the fluid flow around the V-shaped cantilever
gives an accurate prediction of the peak frequency and quality factor.
 
\section{Conclusions}

We have shown that the thermodynamic approach is a versatile and powerful method 
for predicting the stochastic dynamics of cantilevers in fluid for the precise conditions 
of experiment including complex geometries and the presence of nearby boundaries. 
Available analytical predictions are for idealized situations including simple geometries 
where the three-dimensional flow near the cantilever tip has been neglected. Although 
this has provided significant insight, many situations of experimental interest are more 
complicated. It is often required to have a quantitative base-line understanding of the 
cantilever dynamics for the precise conditions of experiment in order to make and interpret 
measurements in novel situations and in the presence of other phenomena of interest.

We emphasize that by using the fluctuation-dissipation theorem a single deterministic 
calculation is sufficient to predict the stochastic behavior for all frequencies. Furthermore,  
the deterministic calculation is computationally inexpensive and does 
not require special computing resources. 

The thermodynamic approach is general in that it can be 
used to compute the stochastic dynamics of any conjugate pair of variables. We have 
shown that the stochastic dynamics that are measured depend upon the choice of 
measurement. This could be exploited in future experiments, for example, to minimize or maximize 
the significance of the higher mode dynamics by choosing to measure tip-deflection 
or tip-angle, respectively.

Our results also suggest that despite the complicated three-dimensional 
nature of the flow field around a V-shaped cantilever, analytical predictions based upon 
a two-dimensional description are surprisingly accurate if the appropriate length 
scales are used.  We anticipate that these findings will be of immediate use as 
the atomic force microscope continues to find further use in liquid environments.

\noindent Acknowledgments: This research was funded by AFOSR 
grant no. FA9550-07-1-0222. Early work on this project was funded 
by a Virginia Tech ASPIRES grant. We would acknowledge many 
useful interactions with Michael Roukes, Mike Cross, and 
Sergey Sekatski.

\end{document}